\documentstyle[twocolumn,prb,aps]{revtex}

\def\*{{\vskip3truemm}}\def\ie{{\it i.e.\ }}\def\eg{{\it e.g.\ }}
\let\lis=\overline\def\V#1{{\mathbf#1}}\let\nota=\small
\def\annota#1{\footnote{\small#1}}\def\etc{{\it etc.}}
\let\0=\noindent\def\media#1{{\langle\,#1\,\rangle}}
\def\Dpr{{\V\partial}}\def\ciao{\end{document}}
\def\W#1{#1_{\kern-3pt\lower6.6truept\hbox to 1.1truemm
{$\widetilde{}$\hfill}}\kern2pt\,}
\let\a=\alpha  \let\d=\delta
 \let\z=\zeta \let\h=\eta
\let\th=\vartheta\let\k=\kappa \let\l=\lambda \let\m=\mu \let\n=\nu
\let\x=\xi \let\p=\pi \let\r=\rho \let\s=\sigma \let\t=\tau
 \let\ch=\chi  
  \let\D=\Delta 
\let\X=\Xi   \let\F=\Phi

\def\bline{\hbox to\hsize}
\newdimen\xshift \newdimen\xwidth \newdimen\yshift
\def\ins#1#2#3{\vbox to0pt{\kern-#2pt \hbox{\kern#1pt #3}\vss}\nointerlineskip}
\def\eqfig#1#2#3#4#5{\par\xwidth=#1pt \xshift=\hsize \advance\xshift
by-\xwidth \divide\xshift by 2\yshift=#2pt \divide\yshift by 2
\bline{\hglue\xshift \vbox to #2pt{\vfil#3 \includegraphics{#4.ps}
}\hfill\raise\yshift\hbox{#5}}}
\def\8{\write13}

\begin{document}

\relax
\preprint{R1/99/2-gg}

\title{Ergodic and chaotic hypotheses: nonequilibrium
ensembles in statistical mechanics and turbulence}

\author{Giovanni Gallavotti}
\address{Fisica, Universit\`a di Roma ``La Sapienza''}

\date{15 may 1999}

\maketitle 
\begin{abstract} The ergodic hypothesis outgrew from the
ancient conception of motion as periodic or quasi periodic. It did
cause a revision of our views of motion, particularly through
Boltzmann and Poincar\'e: we discuss how Boltmann's conception of
motion is still very modern and how it can provide ideas and methods
to study the problem of nonequilibrium in mechanics and in
fluids. This leads to the chaotic hypothesis, a recent interpretation
of a very ambitious principle conceived by D. Ruelle: it is a possible
extension of the ergodic hypothesis and it implies general
parameterless relations. Together with further ideas, it appears to be
consistent with some recent experiments as we discuss here.
\end{abstract} 

\begin{section}{Ergodic hypothesis}
\narrowtext

Since Galileo's {\it ``Philosophy is written in this great book which
is continuously open before our eyes''}, 377 years ago [Ga65], p.38,
we deciphered a few more pages of the great book, beyond the ones that
had already been read in the $~3000$ previous years. The substantial
conceptual identity between the problems met in the theoretical study
of physical phenomena is absolutely unexpected and surprising, whether
one studies equilibrium statistical mechanics, or quantum field
theory, or solid state physics, or celestial mechanics, harmonic
analysis, elasticity, general relativity or fluid mechanics and chaos
in turbulence.  I discuss here a few aspects of the developments of
the theory of chaos as a paradigm of the stability of our processes of
understanding natural science.

In the Renaissance Copernicus started anew the theoretical foundations
of astronomy: since Ptolemy the technical ability to understand the
``world'', \ie the motion of the planets, had been essentially
lost. With Copernicus comes the rediscovery of the technical meaning
of the Greek conception of motion as generated by many {\it uniform
circular motions}: in his youth he undertakes to improve Ptolemy's
great work by restoring the simplicity of the Aristotelian conception
that he thought Ptolemy had betrayed.\annota{``{\it Nevertheless, what
Ptolemy and several others legated to us about such questions,
although mathematically acceptable, did not seem not to give rise to
doubts and difficulties}'' ...  ``{\it So that such an explanation did
not seem sufficiently complete nor sufficiently conform to a rational
criterion}'' ... ``{\it Having realized this, I often meditated
whether, by chance, it would be possible to find a more rational
system of circles with which it would be possible to explain every
apparent diversity; circles, of course, moved on themselves with a
uniform motion''}, [Co30], p.108.}

At the end of his work he left us a system of the world apparently
more orderly than that of Ptolemy, if not more precise. I say
apparently because it seems to me that Ptolemy's {\it Almagest} is
more an astronomical almanac than a book in which the theory of
celestial motions is discussed. It would be difficult, if not
impossible, to extract from the modern {\sl Astronautical Almanac},
[AA89], informations about the three body problem: it is not
impossible that we simply ignore, as Copernicus did, the theory at the
base of the compilation of the Almagest whose ``{\it explanation did
not seem sufficiently complete nor sufficiently conform to a rational
criterion}'', [Co30] p.108. Often ancient science has been
misinterpreted because its original purpose had been forgotten or had
become corrupted, [Ru98].

I would say that Copernicus' contribution, far greater than ``just'' setting
the Earth aside by the second postulate of his {\it Commentariolus},
[Co30], was to show how a consistent ``system of the world'' could be
developed from scratch (\ie from raw observations): the method that he
followed generated a systematic rethinking of the structure of the
``world'' (in this case the system of the planets) which led or at
least guided the works of Galileo, Kepler and many others until the
Newtonian synthesis was achieved, whose all encompassing power is
expressed in the work of Laplace. With Laplace's work the Greek
conception of motion had again become very clear and understood. With
the addition of methods to {\it deduce} cycles and epicycles starting
from very simple first principles (the law of gravitation): the
enthusiasm of the new scientists was so overwhelming that the
classical names became (and remain) obsolete with, for instance, the
epicycles and deferents becoming the austere Fourier modes that could
be read from the tables of Le Verrier.

It was at this moment of triumph of the orderly and simple motion by
cycles, deferents and epicycles, with the inebriating sense of power
that must have been felt when (for instance) the periods of the Moon
became (easily) computable from first principles, that the atomic
hypothesis started being investigated beyond its first
steps. Boltzmann's attempts to derive thermodynamics from mechanics
and the atomic hypothesis really began undermining the conceptions of
motions that all scientists had maintained for millennia, with
relatively minor changes. At the beginning all seemed to indicate just
a new addition to the old Aristotelian views: in his early papers
Boltzmann is perhaps somewhat uneasy with nonperiodic motions: he
prefers to think of a nonperiodic motion as of ``a periodic motion
with infinite period'', [Bo66].\annota{ See p. 30: ``{\it ... das man
die bahnen, falls sie in keiner endlichen Zeit geschlossen sind, doch
in einer unendlichen Zeit als geschlossen ansehen darf.}'' (!) In fact
if one looks at the context in which the above statement is made one
realizes that it is completely justified and that Boltzmann never
really changed his mind in the later work.}

And Boltzmann's mechanical definition of entropy addresses the
conflicting notions of ``measure of disorder'' on the one hand, and of
a property of systems whose motions are very ordered, necessarily
``periodic''. His fundamental work, [Bo84], where he lays down the
theory of statistical ensembles, in a form that is astoundingly modern
and almost identical to the one we use today, associates entropy with
the mechanical properties of periodic motions: he even starts the
paper by showing that one can associate a function ``with the
properties of the entropy'' to the motion of a Saturn ring, regarded
as a rigidly rotating circle (which is so unusual an example that it
is likely to be the reason why such a fundamental paper has been
little noted).

But, as Boltzmann himself had to argue against the objections of
Zermelo, [Ce99], nothing could be less ordered that the motions to which
he was trying to attach a quantity to be identified with entropy, which
also appeared to play a rather different role in his theory of approach
to equilibrium via the Boltzmann equation.

These were the years when Poincar\'e had noted the recurrence theorem,
from which some wanted to derive the proof of the alleged
inconsistency of the atomic hypothesis, [Ce99], viewing matter as an
assembly of particles obeying Newton's equations, because of its
conflict with macroscopic thermodynamics. At the same time Poincar\'e
had for the first time given incontrovertible evidence that planetary
motions {\it could not always be explained} in terms of cycles and
epicycles (as Laplace theory of the world hinted): I refer here to his
theorem, [Po87], ``of nonintegrability'' of the three body
problem. The inescapable consequence, of which Poincar\'e was well
aware, was that not all motions could be quasi periodic, \ie
compositions of circular motions.

Nevertheless one of the achievements of Boltzmann was the {\it heat
theorem}: to a system endowed only with periodic motions, a {\it
monocyclic system} after Helmoltz, [Bo84], one could associate {\it
mechanical} quantities, that could be given a name familiar from
macroscopic physics, like ``temperature'' $T$, ``energy'' $U$,
``volume'' $V$, ``pressure'' $p$, so that by changing infinitesimally
the parameters describing the system the consequent changes $dU$ and
$dV$ of $U$ and $V$ would be such that
\begin{equation}\frac{dU+ p dV}T= \ {\rm exact\
differential}\label{(1.1)}\end{equation}
which can be used to define entropy as the integral of the exact
differential: and eq. (\ref{(1.1)}) is the analytic form of the second
law of thermodynamics in equilibrium. In the late work of Boltzmann,
[Bo84], where it is proved in maximum generality, this theorem appears
as a consequence of his {\it ergodic hypothesis}: an hypothesis that
has a double nature. On the one hand it is usually interpreted as
saying that the motion (in phase space) is rather random; on the other
hand it rests on an essential idea of Boltzmann, that in fact phase
space can be regarded as discrete (basically because we cannot suppose
that the world is a continuum: see ([Bo74]), p. 169, \*

{\it Therefore if we wish to get a picture of the continuum in words,
we first have to imagine a large, but finite number of particles with
certain properties and investigate the behavior of the ensemble of
such particles. Certain properties of the ensemble may approach a
definite limit as we allow the number of particles ever more to
increase and their size ever more to decrease. Of these properties one
can then assert that they apply to a continuum, and in my opinion this
is the only non-contradictory definition of a continuum with certain
properties.}
\*

A similar view was held for the phase space in which atoms are
described, [Ce99], [Ga95].

If phase space is regarded as discrete then every motion is a
permutation of its discrete points, called ``cells'', hence it {\it
must} be periodic and it is then reasonable that it is just a one
cycle permutation of the cells on the energy surface. Thus Boltzmann
hypothesizes that motion, viewed as a permutation of cells with the
same energy, has one cycle, \ie that every cell visits all the others
before returning to itself.

Hence we see the duality mentioned above: to derive thermodynamics we
assume that the motion is periodic, but at the same time such
that the motion of the system is so irregular to fill the whole energy
surface. It is not surprising that many scientists were shocked by
arguments and theories built on apparently conflicting assumptions:
they disregarded Boltzmann's discrete approach and, identifying cells
with points of a {\it continuous} energy surface, pointed out the
mathematical inconsistency of the ergodic hypothesis strictly
interpreted as saying that a point representing the system in phase
space wanders around passing eventually through {\it every} point of
the energy surface (quite absurd in general, indeed), see p. 22 and
notes 98, 99 at p. 90 in [EE11].

A key point that is often overlooked is that the relation
(\ref{(1.1)}) is a property that holds for arbitrary mechanical
systems of identical particles no matter how small or large they are,
[Bo84] and Ch. I of [Ga99b]. Only for assemblies of atoms that can be
considered to form a macroscopic system the quantities $U,T,V,p$
acquire the interpretation that is suggested by our familiarity with
their names, and therefore only for such systems eq. (\ref{(1.1)}) can
be regarded as the second law of equilibrium thermodynamics, [Bo84]. A
``trivial'' general identity can then be interpreted as a very
important law of nature, [Ga99b].

The long discussions on the matter, led initially by Boltzmann who
began to explain to ears unwilling to listen why there was no
contradiction in his discoveries, continued until today with every new
generation bringing up the same old objections against Boltzmann's
theories often (though not always, of course) still refusing to listen
to the explanations (for a modern discussion of Boltmann's views see
[Le93], [Ga95]).

But one can say that after Boltzmann there was no substantial
progress, at least no better understanding was gained on the
foundations of Statistical Mechanics other than the theorem of Lanford
proving Grad's conjecture on the possibility of deriving rigorously
the Boltzmann equation from a microscopically reversible dynamics,
[La74] (a result which unfortunately does not seem to be as well known
as it should).

The achievement of Boltzmann was to have proposed a {\it general
assumption} from which one could derive the prescription for studying
properties of large assemblies of particles (I refer here to the ergodic
hypothesis and, [Bo84], to the microcanonical and canonical ensembles
theory) and that had immediately proved fruitful through its prediction
of the second law, (\ref{(1.1)}). This remained an isolated landmark while
interest concentrated on the derivation of further consequences of the
new theory: namely to understand phase transitions and their critical
points, or the basic quantum statistical phenomena: black body
radiation, superconductivity or superfluidity are, perhaps, the clearest
examples.

Also the parallel efforts to understand phenomena out of equilibrium
were far less successful. Yet in a sense such phenomena too {\it must
be understood}, not only because of their obvious interest for the
applications, which most often deal with systems in stationary
nonequilibrium states, like a turbulent flow of a liquid in a pipe or
a stationary current kept in a circuit by an electromotive force, but
also because their understanding promises to bring light on the
mentioned duality between {\it orderly motions}, \ie periodic or quasi
periodic, and {\it chaotic motions}, as we now call motions that are
{\it neither} periodic {\it nor} quasi periodic.  \*

It is not until the 1960's, under the powerful solicitation of new
experimental techniques and the rapid growth of digital computers,
that the problem began to be attacked. Existence of chaotic motions
became known and obvious even to those who had no familiarity with the
work of Poincar\'e and with the results, [Si77], of Hopf, Birkhoff,
Anosov and more recently of Kolmogorov, Sinai and many others that
developed them further.

Works on chaotic motions started to accumulate until their number
really ``exploded'' in the 1970's and it continued to grow rapidly,
since. The goal of the research, or at least one of the main goals,
was to understand how to classify chaotic phenomena whose existence
had become known and visible to the (scientific and not scientific)
general public which seemed quite surprised for not having noted them
before. Perhaps the main aim was to find out whether there was any
extension to nonequilibrium systems of the statistical ensembles that
were at the basis of the applications of equilibrium statistical
mechanics.

The problem has two aspects which initially seemed uncorrelated:
indeed chaotic motions arise both in many particles systems typical of
statistical mechanics and in fluids (and in other fields not
considered here, for lack of space).

It is in the theory of fluids that the last attempt to an Aristotelian
interpretation of motion had survived to these days. The book of
Landau and Lifschitz, [LL71], presents a remarkable theory of fluid
turbulence based on quasi periodic motions: basically a fluid in a
container of fixed geometry put in motion by external constant (non
conservative) forces would settle in a stationary state which would
look at first, under weak forcing, static (``laminar''), then periodic
(in Greek terms it would be described by ``one epicycle'') then quasi
periodic with two periods (in Greek terms it would be described by
``two epicycles'') then periodic with three periods (``three
epicycles'') and so on until the number of epicycles had grown so
large and, hence, the motion so complex to deserve the name of
``turbulent''.\annota{Unfortunately the quoted chapter on turbulence
has been removed from the more recent editions of the book and
replaced by a chapter based on the new ideas: a choice perhaps useful
from the commercial viewpoint but quite criticizable from a
philological viewpoint. Of course keeping the original version and
adding the new one as a comment to it would have been more expensive:
a saving that might generate a lot of work a thousand years from now
and that continues a long tradition which makes us wonder even what
Euclid really wrote and what might have been added or changed later.}
\*

Through the work of Lorenz, [Lo63], and of Ruelle--Takens, [RT71a], it
became clear that the quasi periodic view of the onset of turbulence
was untenable: a conclusion which also several Russian scientists
had apparently reached, [RT71b], independently.

The works making use of the new point of view stem also, and
mainly, from the innovative ideas that Ruelle later wrote or simply
exposed in lectures. There he developed and strongly stressed
that the mathematical theory of dynamical systems, as developed in
this century, would be relevant and in fact it would be the natural
framework for the understanding of chaotic phenomena.

The impact on experimental works was profound. Already the very fact,
[RT71a], that a study of the onset of turbulence could be physically
interesting had been new at the time (the 1960's and early 1970's).
And one can say that after the first checks were performed, some by
skeptical experimentalists, and produced the expected results a stage
had been achieved in which the ``onset of turbulence'' was so well
understood that experiments dedicated to check the ``Ruelle--Takens''
ideas on the onset of turbulence were no longer worth being performed
because one would know what the result would be. 
\*

In this respect, before proceeding to the (developed) turbulence
problems we stress that there remains still a lot to be done: the
phenomena appearing at the onset of turbulence are in a sense too fine
and detailed, and besides telling us that motions can be far more
complex than one would have imagined a priori they give us little
perspective on the theory of developed turbulence, admittedly more
difficult.

Understanding the onset of turbulence is perhaps analogous to
understanding the atomic system and classifying the spectra. The
variety of the atomic spectra is enormous and its classification led
quite naturally to quantum theory: but in itself it is of
little help in understanding the mechanical properties of gases or of
conducting metals, for instance.

Likewise we should expect that the analysis of the onset of turbulence
will eventually lead to a more fundamental understanding of how the
basic chaotic motions (that appear in a, so far, imperscrutable way at
the onset of turbulence) are in fact predictable on the basis of some
general theory: we have many experiments and a wide corpus of
phenomena that have been studied and recorded and the situation is
similar to the one at the beginning of the century with the atomic
observations. {\it We see a few types} of ``bifurcations'', \ie
changes in the stationary behavior of systems, that develop in many
different systems, as the strength of forcing is increased, but we do
not know how to predict the order in which the different bifurcations
arise and why they do so.

In a way it is deceiving that this understanding has not yet been
achieved: this is certainly a goal that we should have in mind and
that perhaps will be attained in a reasonable time in view of its
practical importance.  But we cannot expect that the solution, much
desired as it is by all, can by itself solve the problems that we
expect to meet when we study the stationary behavior of a {\it
strongly turbulent} fluid or a gas of particles out of
equilibrium. Much as understanding the two body problem gives us
little {\it direct} information on the behavior of assemblies of
$~10^{19}$ particles (corresponding to $1\,cm^3$ of Hydrogen in normal
conditions).  \*

In the light of the above considerations it is important to note that
Ruelle's view, besides reviving the interest in Lorenz' work which had
not been appreciated as it should have been, was noticed by physicists
and mathematicians alike, and had a strong impact, because it was {\it
general and ambitious in scope} being aimed at understanding from a
fundamental viewpoint a fundamental problem.

In 1973 he proposed that the probability distributions describing
turbulence be what are now called ``Sinai-Ruelle-Bowen''
distributions. This was developed in a sequence of many technical papers
and conferences and written explicitly only later in 1978, [Ru78], see
also [Ru99]. It had impact mostly on numerical works, but it proposes a
fundamental solution to the above outstanding theoretical question: what
is the analog of the Boltzmann--Gibbs distribution in non equilibrium
statistical mechanics? his answer is a general one valid for chaotic
systems, be them gases of atoms described by Newton's laws or fluids
described by Navier--Stokes equations (or other fluid dynamics
equations).

It is not simple to derive predictions from the new principle which,
in a sense, is really a natural extension of Boltzmann's ergodic
hypothesis. Systems under nonconservative forcing must be subject also
to forces that take out the energy provided to the system by the
external forces: otherwise a stationary state cannot be
reached. Assuming that the forces are deterministic the equations of
motion {\it must be dissipative}: this means that the divergence of
the equations must have a negative time average and, therefore, the
statistics of the stationary state will be concentrated on a set of
{\it zero volume} in phase space. In other words the motions will evolve
towards an {\it attractor} which has zero volume.

It is precisely the fact that the attractor has zero volume that makes
it difficult to study it: we are not used to think that such singular
objects may have a physical relevance.

However from a point of view similar to the discrete viewpoint of
Boltzmann such a situation is {\it not really different} from that of a
system in equilibrium. One has to think of the attractor as a
discrete set of points and of the dynamics as a permutation of them
which has only one cycle. Then of course the stationary state will be
{\it identified with the uniform distribution on the attractor}, giving
equal probability to each of its points.

The difficulty is that we do not know where the attractor is.  In
equilibrium the problem did not arise: because the attractor was
simply the entire surface of constant energy.

In the next section I discuss from a more technical viewpoint the
meaning of the principle arguing that it is a natural and deep
extension of the ergodic hypothesis. I will then analyze the
potentialities of the hypothesis that, in a form slightly broader than
the original, I will call {\it chaotic hypothesis}, following
[GC95], by showing (\S4) that it is capable of yielding general
universal results, \ie ``parameterless laws'', and perhaps even to
shed some light (\S3) on the very controversial question: ``what is
the proper extension of the notion of entropy to nonequilibrium
systems?''. In \S5 I discuss the notion of dynamical statistical
ensembles and the possibility of equivalence between time reversible
and time irreversible dynamics in ``large'' systems. In \S6 I
attempt at an application of the ideas in \S5 to the interpretation of
an experiment and in \S7 I collect a few conclusions and comments. I
try to avoid technicalities, yet I try not to hide the problems, which
means that for instance in \S6 I must refer to some equations that
might be not familiar to the reader. The references should help the
readers interested in a more technical understanding.
\end{section} 

\begin{section}{Chaotic systems from a technical viewpoint.} 

Observing motion at steps over short time intervals is a natural way to
study time evolution: indeed this is what is normally done in
experiments where observations are always timed in coincidence with some
event of special interest (\eg with the passage of a clock arm though the
``12'' position or, more wisely, with the realization of some event
characteristic of the phenomenon under investigation, \eg a collision
between two particles if the system is a system of balls in a vessel).

Any simulation represents phase space as discrete and time evolution
as a map $S$ over the discrete points. It is assumed that the small
size of the cells (of the order of the machine precision) is so small
that errors, due to the fact that the size is not strictly zero, do
manifest themselves over time scales that are negligible with respect
to the ones over which the phenomena of interest naturally take
place.  Following Boltzmann we shall take the same viewpoint even when
considering real (\ie not simulated) systems and we shall suppose the
phase space to consist of a discrete set of points, also called
``cells''.
\*

We consider a ``chaotic system'' under external forcing and subject to
suitable ``thermostats'', \ie forces that forbid unlimited
transformation into unreleased ``heat'' (kinetic energy) of the work
performed on the system by the external forces, so that the system
{\it can} reach a stationary state (\ie does not ``boil out of
sight''). The evolution will then be described by a map $S$ of the
discrete phase space.

The map $S$ will not, however, be in general a permutation of
cells. Because the effect of the thermostating forces will be that
dynamics will be effectively dissipative, \ie the divergence of the
equations of motion will not vanish and will have a negative average
(unless the system is conservative and therefore the thermostating
forces vanish). Hence a small ball $U$ in phase space will evolve in time
becoming a set $S^T U$ at time $T$ which has a much smaller volume
than the original $U$ and in fact has a volume that tends to $0$ as
$T\to\infty$, usually exponentially fast.

As mentioned in \S1 we must understand better the structure of the
attractor and the motion on it. To visualize the attractor we imagine,
for simplicity of exposition, that the evolution map $S$ has at least
one fixed point $O$ (\ie a configuration in phase space that, observed
with the timing that defines $S$, reproduces itself because it
generates a motion whose period is exactly that of the timing): this
turns out to be not really an assumption\annota{Because a chaotic
system will always have a lot, [Sm67], of periodic orbits and a a
periodic orbit can play the same role plaid here by the fixed point.}
but it is useful, at first, for expository purposes as it eliminates a
number of uninteresting technical steps).
\*

We get a good approximation of the attractor simply by identifying it
with the set $S^TU$ into which a small ball around $O$ evolves in a
large time $T$. The ball will expand strongly along the unstable
manifold of the fixed point $O$ and it will strongly contract along
the stable manifold (as we shall see the point $O$ has to be
hyperbolic, ``together with all the others'', for the picture to be
consistent).

If $T$ is large the image $S^TU$ so obtained will be a very wide and
thin layer of points around a wide portion of the unstable manifold of
$O$, and this layer will be a good approximation of the attractor. The
assumption that the system is thermostated is translated technically
into the property that the region of phase space that the trajectories
starting in $U$ will visit is finite: therefore the unstable manifold
will necessarily ``wound around'' in meanders and the layers will
locally look as stacks of surfaces thinly coated by the points of
$S^TU$.

The layers however will in general {\it not be equispaced} (not even
very near a given point: the case of a conservative system being
essentially the only notable exception) so that a cross section of the
stack of layers will usually remind us more of a ``Cantor set'' than
of a pile of sheets. Furthermore the width of the layer will not be
constant along it but it will change from point to point because the
expansivitity of the unstable manifold is not uniform, in general (not
even in the conservative cases).

We now think phase space, hence also the region $S^T U$, as consisting
of very tiny cells.  The picture of the dynamics will then be the
following: cells which are outside the region $S^TU$ will eventually
evolve into cells inside $S^TU$ while cells inside $S^TU$ will be
simply permuted between themselves.\annota{If we take $T'\gg T$ and
consider instead as a model for the attractor the set $S^{T'}U$ the
picture is unchanged. Even though the volume of the region $S^{T'}U$
is much smaller than that of the region $S^TU$ because the layer
around $S^{T'}U$ is much thinner than that around $S^TU$ while the
portion of surface of the unstable manifold of the fixed point $O$
coated by $S^{T'}U$ is much wider than that covered by $S^TU$. Of
course since contraction prevails over expansion the thinning of the
layer far outweighs the widening of the surface coated (by
$S^{T'}U$).}

One should think that the region $S^TU$ is invariant under the
application of $S$ in spite of the fact that this apparently contradicts
the invertibility of the evolution $S$ (when $S$ is generated by a
differential equation). The point being that a dynamics that evolves
contracting phase space cannot be represented as an invertible
permutation of cells: so that we cannot any more regard
the map $S$ as strictly invertible once we decide to approximate it with
a map on a discrete space. By replacing $T$ with $T'\gg T$ the
approximation improves but it can become exact only when we reduce the
cells size to points, \ie when we use a continuum representation of
phase space.

Therefore it is clear that in order that the above picture be
rigorously correct (\ie in order to be able to estimate the errors
made in the predictions derived by assuming it correct) one needs
assumptions. It is interesting that the ``only'' assumptions needed
are that the continuum system be ``chaotic'' in the sense that pairs
of points initially very close get far apart at a constant rate as
time evolves (\ie exponentially fast) with the exception of very
special pairs. 

More precisely if we follow the motion of a point $x$ in phase space so
that it looks to us as a fixed point $\lis x$, then the action of the
map on the nearby points generates a motion relative to $\lis x$ like that
of a map having $\lis x$ as a hyperbolic fixed point with nontrivial
Lyapunov exponents (\ie with exponents uniformly, in $x$, away from $1$,
some of which larger and some smaller than $1$).

One says that {\it in a chaotic system instability occurs at every
point in phase space}, [Si79], and that a system is chaotic if the
attractors\annota{In general there can be several, as a system can
consist of several non interacting systems represented by points of
sets located in different regions of phase space.} have the above
property whose formal mathematical definition can be found in [Sm67]
and is known as the ``axiom A property'': it is the formal
mathematical structure behind the simplest chaotic systems.

{\it Finally the discrete evolution on the attractor should be ``ergodic'',
\ie the permutation of the cells in $S^TU$ should be a one cycle
permutation.}

The latter property remarkably follows from the chao\-ti\-ci\-ty assumption
and the principle of Ruelle, that I interpret as ``empirical chaoticity
manifests itself in the technical sense that one can suppose, for the
purposes of studying the statistical properties of systems out of
equilibrium, that they have the mathematical structure of systems
with axiom A attractors'' has, therefore, conceptually very strong
consequences.

\end{section}

\begin{section}{The chaotic principle. Entropy and thermostats.}

The principle discussed in the previous sections was originally
formulated for models of (developed) fluid turbulence: here I shall
discuss a slightly different form of it, introduced and applied in
[GC95]
\*

{\it Chaotic Hypothesis: A chaotic system can be regarded, on its
attractor and for the purpose of evaluating statistical properties of
its stationary states, as a transitive Anosov system.}
\*

\0This is stronger than Ruelle's formulation because it replaces axiom A
system by Anosov system (a transitive Anosov system can be thought of
as a dynamical system on a smooth surface which is also an axiom A
attractor). Intuitively one is saying that the attracting set is a
smooth surface rather than a generic closed set.

Implicitly the hypothesis claims that ``fractality'' of the attractor
must be irrelevant in systems with $10^{19}$ or with just many degrees
of freedom.

The hypothesis allows us immediately to say that the stationary state
is uniquely determined and therefore we are in a position similar to
the one in equilibrium where also, by the ergodic hypothesis, the
statistics of the equilibrium state was uniquely determined to be the
microcanonical one. And if applied to a system in equilibrium (\ie to
a system of particles subject to conservative forces) it gives us
again that the statistics of the motions is precisely the
microcanonical one.

In other words the chaotic hypothesis is a strict extension of the
ergodic hypothesis and it provides us with a formal expression
(``uniform distribution on the attractor'')\annota{Or, as one says
more technically but equivalently for Axiom A attractors,
``distribution absolutely continuous along the unstable manifolds''.}
for the analogue of the microcanonical ensemble in systems out of
equilibrium but stationary. The new distribution is called the SRB
distribution of the system.

As discussed in \S2 the hypothesis amounts to assuming that the
motion is periodic. Hence the dualism between periodic and chaotic
motions persists in the same sense as in the case of Boltzmann's
equilibrium theory. And, as in that case, one should not confuse the
periodic motion on the attractor with the periodic motions of
``Aristotelian'' nature: the latter are motions with short, observable
periods, on the same time scale of the observation times. This is also
the case in Laplace's celestial mechanics and in the Landau--Lifshitz
theory of turbulence. The periods of the motions involved in the
chaotic hypothesis are unimaginably larger; in the case of a model of
a gas (in equilibrium or in a stationary nonequilibrium state) the
period is estimated by Boltzmann's well known estimate to be ``about''
$10^{10^{19}}$ {\it ages of the universe}, [Bo95] p. 444.
\*

We begin to explore the consequences of the chaotic hypothesis by
looking at the notion of {\it entropy}. To fix the ideas we imagine a
Hamiltonian system of $N$ particles in a box ${{\cal B}}$ which is forced by a
constant external force and is in contact with $s$ heat reservoirs
${{\cal R}}_k,\, k=1,\ldots,s$. We assume that the box opposite sides in the
direction parallel to the force field $\V E$ are identified and that
inside the box there are fixed scatterers (\eg on a regular array),
enough so that there is no stright path parallel to $\V E$ which does
not hit a scatterer. A symbolic illustration of the situation is in the
following picture with two reservoirs


\eqfig{187.5}{97.5}{
\ins{9}{75}{$\scriptstyle {{\cal R}}_1$}
\ins{120}{75}{$\scriptstyle {{\cal R}}_2$}
\ins{-15}{22.5}{$\scriptstyle {{\cal B}}$}
\ins{37.5}{78.75}{$\scriptstyle N_2$}
\ins{142.5}{78.75}{$\scriptstyle N_2$}
\ins{105}{37.5}{$\buildrel \scriptstyle \V E\over \longrightarrow$}
\ins{75}{37.5}{$\scriptstyle N$}}{tmpuvw1}{}
\*
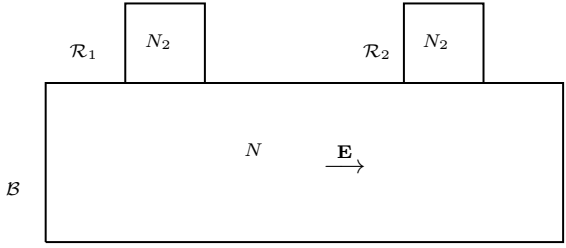
\0{\nota{}Fig.1:\it The scatterers in the box ${{\cal B}}$ are not drawn; 
the particles in the reservoirs interact between themselves and with
those of the system. The opposite sides perpendicular to $\V E$ are
identified.}
\*

This means that the equations of motion are
\begin{equation}m {\V {{\ddot  x}}}_i={\V f}_i+\V E+ {\V\th_i},\qquad
\V f_i=\sum_{j\ne i} \V f(\V x_i-\V y_j)\label{(3.1)}\end{equation}
where $\F f(\X x-\V y)$ is the (conservative) force that a particle at
$\V y$ exerts over one at $\V x$, $\V\th_i(t)$ are the forces due to
the thermostats and $m$ is the mass.

The particular form of the thermostating forces should be, to a large
extent, {\it irrelevant}. Therefore we make the following model for the
thermostats. Each of the $s$ thermostats is regarded as an assembly of
$N_k,\, k=1,2,\ldots,s$ particles which are kept at constant
temperature:
\begin{equation}\V\th^{(k)}_i=\sum_{j=1}^{N_k} \V {{\tilde f}}^{(k)}(\V x_i-\V
y_i^{(k)})-\a\,\V{{\dot x}}_i\label{(3.2)}\end{equation}
where $\V {{\tilde f}}^{(k)}(\V x-\V y^{(k)})$ is the (conservative)
force that the thermostat particle at $\V y$ exercises over the system
particle at $\V x$, while the particles of the $k$--th reservoir
satisfy the equation
\begin{equation}m \V{{\ddot y}}_i=\V f^{(k)}_i-\a^{(k)}\, \V{{\dot
y}}^{(k)}_i\label{(3.3)}\end{equation}
where $\V f_i^{(k)}$ is the (conservative) force that the particle at
$\V y_i^{(k)}$ feels from the other particles of the $k$--th
thermostat or from the system particles.

The multipliers $\a,\a^{(k)}$ are so defined that the temperatures of
the system and that of each reservoir is fixed in the sense that, if
$k_B$ denotes Boltzmann's constant,
\begin{eqnarray}
&\frac1{N_k}\sum_{j=1}^{N_k} \frac{m}{2} ({\V{\dot
y}}^{(k)}_j)^2=\frac32 k_B T_k,\nonumber\\ 
&\frac1N \sum_{j=1}^N
\frac{m}2 ({\V{\dot x}}^{(k)}_j)^2=\frac32 k_B
T\label{(3.4)}\end{eqnarray}
are exactly constant along the motions. The model is obtained by
requiring that the constraints (\ref{(3.4)}) are imposed by exerting a
force that satisfies the {\it principle of minimum constraint} of
Gauss (see appendix in [Ga96a]), just to mention a possible model of a
thermostat widely used in applications then the multipliers take the
values
\begin{eqnarray}
\a^{(k)}=&\frac{\sum_j^{N_k} 
\V f^{(k)}_j\cdot\V {{\dot y}}^{(k)}_j}{\sum_j ({\V {{\dot y}}^{(k)}_j})^2}=
\frac{\dot Q_k}{3 k_B T_k}\nonumber\\
\a=&\frac{\sum_j^{N} (\V f^{tot}_j+\V E)\cdot\V {{\dot x}}_j}
{\sum_j {\V {{\dot x}}_j}^2} =\frac{\dot Q}{3 k_B T}+ \frac{\V E\cdot\V
J}{3 k_BT}\label{(3.5)}\end{eqnarray}
where $\V J=\sum_j^N \V {{\dot x}}_i$ is the ``{\it current}'' and $\V
f^{tot}_i=\V f_i+ \sum_k\sum_j \V{{\tilde f}}^{(k)}(\V x_i-\V
y_j^{(k)})$ is the total force acting on the $i$--th particle.

If we compute the divergence of the equations of motion in the phase
space coordinates $(\V p,\V q)$ with $\V p_i=m
\V{{\dot x}}_i$, $\V p^{(k)}_i=m \V{{\dot y}}^{(k)}_i$ we get, as
noted in eq. (3.4) of [Ga96a],

\begin{equation}
\sum_{k=1}^s \frac{\dot Q_k}{k_BT_k}+\frac{\dot Q}{k_B T}+\frac{{\V
J\cdot \V E}}{k_B T}\label{(3.6)}\end{equation}
up to corrections of order $N_k^{-1}$ and $N^{-1}$.\annota{The exact
value is obtained by multiplying the $k$--th term in the sums by
$1-\frac1{3N_k}$ and the other terms by $1-\frac1{3N}$.}  If there is
no external field $\V E$ or if the temperature $T$ is not kept fixed,
we only get
\begin{equation}
\sum_{k=1}^s \frac{\dot Q_k}{k_B T_k}\label{(3.7)}\end{equation}
still up to corrections of order $N_k^{-1}$ and $N^{-1}$.

The quantities $-\dot Q_k$, $-\dot Q-\V E\cdot \V J$ represent the
work done over the system (including the thermostats) to keep the
temperatures fixed: this means that if the system is in a stationary
state the same quantities changed in sign must represent the heat that
the thermostats cede to ``the outside'' in order to function as
such. So (\ref{(3.6)}) or (\ref{(3.7)}) represent the {\it rate of increase of
the entropy} of the ``Universe'' (in the sense of thermodynamics).

It is gratifying that, as proved in wider generality by Ruelle,
[Ru96], a system verifying the chaotic hypothesis must necessarily
satisfy the inequality

\begin{equation}
\media{\, \sum_{k=1}^s \frac{\dot Q_k}{k_BT_k}+\frac{\dot Q}{k_B
T}+\frac{{\V J\cdot \V E}}{k_B T}\,}\ge0\label{(3.8)}\end{equation}
where $\media{\cdot}$ denotes time average over the stationary state.
One can check that the contribution to the average due to the internal
forces between pairs of particles in ${{\cal B}}$ {\it vanishes}, therefore
$\media{\dot Q}$ receives contributions only from the forces exerted
by the thermostats on the system.

Hence from the above example, which is in fact very general, the
entropy creation rate when the system is in the phase space point $x$
should be defined {\it in general} and in deterministic, finite,
thermostated systems to be

\begin{equation}
\s(x)=-{\rm\, div\,} F(x)\label{(3.9)}\end{equation}
where $x$ denotes a phase space point describing the microscopic state
of the system and its evolution is given by the differential equation
$\dot x=F(x)$ for some vector field $F$.\annota{One should note,
however, that in general switching on thermostating forces does not
necessarily imply that the system will reach a stationary state: for
instance in the above example with a field $\V E$ but no thermostat
acting on the bulk of the system (\ie without fixing the bulk
temperature $T$) it is by no means clear that the system will reach a
stationary state: in fact the energy exchanges with the thermostats
could be so weak that the work done by the field $\V E$ could
accumulate in the form of an ever increasing kinetic energy of the
particles in the container. We need a nonobvious (if at all true) \ap
estimate of the energy of the bulk which tells us that it will stay
bounded uniformly in time. The result will strongly depend on the
nature of the forces between thermostats particles and system
particles $\V {{\tilde f}}^{(k)}$ and on the interparticle forces.}

It appears therefore reasonable (or {\it it may} appear reasonable) to
set the following definition {\it in general}:
\*

{\it Definition: In a finite deterministic system the instantaneous
entropy creation rate is identified with the divergence of the
equations of motion in phase space evaluated at the point that
describes the system at that instant.}
\*

A strong argument in favor of this ambitious definition is the
following, [An82]. Suppose that a finite system is in a equilibrium
state at some energy $U$ and specific volume $v$. At time $t=0$ the
equations of motion are changed because the system is put in contact
with heat reservoirs and subject to certain external forces whereby it
undergoes an evolution at the end of which, at time $t=+\infty$, the
system is {\it again} governed by Hamiltonian equations and settles
into a new equilibrium state. If $\r_0$ is the density in phase space
of the distribution representing the initial state, $\r_t$ the density
of the state at intermediate time $t$ and $\r_\infty$ is the density
over phase space of the final state then we can study the evolution of
$S(t)=-\int \r_t\log\r_t \,dpdq$ which evolves from $S_0=-\int
\r_0\log \r_0 \,dpdq$ to $S_\infty=-\int \r_\infty\log\r_\infty
\,dpdq$ and one checks that:

\begin{eqnarray}
S_\infty-S_0&=&\int_0^\infty \frac{d}{dt} S(t)=\int_0^\infty dt\,\int
\r_t\, {\rm div\,} F \, d\V p\,d\V q=\nonumber\\ 
&=&\int_0^\infty
\media{{\rm div}\, F}(t)\,dt\hfill\label{(3.10)}\end{eqnarray}
so that we see that also in this case $\media{{\rm div}\, F}(t)$ can be
interpreted as (average) entropy creation rate, [An82].

An argument against the above definition is that it does not seem to
be correct in systems in which the thermostats are modeled by infinite
systems initially in equilibrium at a given temperature and
interacting with the particles of the system that is thermostated: in
such systems there will be a flow of heat at infinity and the above
considerations fail to be applicable, in a fundamental way, as shown
in [EPR98].  However I see no arguments against the definition when
one uses finite thermostats and finite systems.

In nonequilibrium statistical mechanics the notion of entropy and of
entropy creation are {\it not} well established. New definitions and
proposals arise continuously. 

Hence a fundamental definition is highly desirable. By fundamental I
mean a definition, like the one above, that should hold for very
general systems in stationary states: {\it and} it should not be
restricted to (stationary) systems close to equilibrium. This means
that it should be defined even in situations where the other
fundamental thermodynamics quantity, the {\it temperature}, may itself
be also in need of a proper definition. And {\it furthermore} it
should be a notion accessible to experimental checks, on numerical
simulations and possibly on real systems.

\end{section}

\begin{section}{Fluctuations and time reversibility.}

The definition of entropy creation rate in \S3 points in the direction
of an adaptation of the very first definitions of Boltzmann and Gibbs
and relies on the recent works on chaotic dynamics, both in the
mathematical domain and in the physical domain.

The theory of the SRB distributions together with Ruelle's proposal
that they may constitute the foundations of a general theory of
chaotic motions, provides us with {\it formal expressions} of the
probability distributions describing stationary states (namely equal
probability of the attractor cells). This is a surprising
achievement\annota{Relying, for a more technical and usable
formulation, on the basic work of Sinai on Markov partitions, [Si77].}
and the hope is that such formal expressions can be used to derive
relations between observable quantities whose values there is no hope
to ever be able to compute via the solution of the equations of motion
(much as it is already the case in equilibrium statistical mechanics).

I have in mind general relations like Boltzmann's heat theorem
$\frac{dU+p dV}T=\ {\rm exact}$, (\ref{(1.1)}), which involves
averages $U,p$ (computed, say, in the canonical ensemble), where $V,T$
are ``parameters'') that we cannot hope ever to compute, but which
nevertheless is a very important, non trivial and useful relation. Are
similar relations possible between dynamical averages in {\it
stationary nonequilibrium states}? after all a great part of
equilibrium statistical mechanics is dedicated to obtaining similar
(if less shiny) relations, from certain $N$--dimensional integrals
(with $N$ very large) representing partition functions, correlation
functions, \etc

Of course such results are difficult: but they might be not
impossible. In situations ``close'' to equilibrium there are, in fact,
{\it classical examples} like Onsager's reciprocal relations and
Green--Kubo's transport coefficients expressions: these are
parameterless relations essentially independent of the model used, as
long as it is time reversible at least at zero forcing (\ie in
equilibrium).

By ``close'' we mean that the relations are properties of derivatives
of average values of suitable observables {\it evaluated at zero
forcing}: one says that they are properties that hold {\it
infinitesimally close} to equilibrium.

A system is said to be time reversible under a time reversal map $I$
if $I$ is an isometry of phase space with $I^2=1$ which anticommutes
with the time evolution $S_t$, (or $S$ if the evolution is represented
by a map), \ie

\begin{equation}IS_t=S_{-t}I,
\qquad{\rm or}\qquad IS=S^{-1}I\label{(4.1)}\end{equation}
where $S_t x$ denotes the solution of the equations of motion at time
$t$ with initial datum $x$. Clearly $S_{t}S_{t'}=S_{t+t'}$.

The thermostat models that are derived, as in \S3, from the Gauss'
principle have the remarkable property of generating {\it time
reversible} equations of motion. The importance and interest of such
models of thermostats has been discovered and stressed by Hoover and
coworkers, [PH92]. This has been an important contribution, requiring
intellectual courage, because it really goes to the heart of the
problem by stressing that one can (and should) study irreversible
phenomena by only using reversible models: {\it microscopic
reversibility has nothing to do with macroscopic irreversibilty}, as
Boltzmann taught us and getting rid of spurious microscopically
irreversible models can only help our understanding.
\*

In a finite deterministic system verifying the chaotic hypothesis
entropy creation rate fluctuations can be conveniently studied in
terms of the {\it average} entropy creation rate $\s_+$, evaluated on
the stationary state under exam and assumed $>0$, and of the {\it
dimensionless} entropy creation rate

\begin{equation}
p=\frac1\t\int_{-\t/2}^{\t/2}\frac{\s(S_tx)}{\s_+}
dt\label{(4.2)}\end{equation}
The quantity $p$ is a function of $x$ (and of course of
$\t$). Therefore if we observe $p$ as time evolves, in a stationary
state, it fluctuates and we call $\p_\t(p)dp$ the probability that it
has a value between $p$ and $p+dp$. On general grounds we expect
that $\p_\t(p)=\exp \t\z(p)+o(\t)$ for $\tau$ large, [Si77]. The
function $\z(p)$ is a suitable model dependent function with a maximum
at $p=1$ (note, in fact, that by the normalization in (\ref{(4.2)})
the infinite time average of $p$ is $1$).

The theorem that one can prove under the only assumption of the
chaotic hypothesis and of revesibility of the dynamics is
\begin{equation}\z(-p)=\z(p)-p\s_+\label{(4.3)}\end{equation}
where $-\infty\le\z(p)< +\infty$.

The (\ref{(4.3)}) is the analytic form of the {\it fluctuation theorem},
[GC95]: it is a {\it parameterless relation, universal} among the
class of systems that are time reversible and transitive. It was first
observed experimentally in a simulation, [ECM93]; it was proved, and
its relation with the structure of the SRB distributions was
established, in [GC95].  See also [Ru99] for a general theory and
[CG99] for some historical comments.

It is a general relation that holds whether the system is at small
forcing field or not, {\it provided} the system remains transitive \ie
the closure of the attractor is the full phase space in the sense that
it is a time reversal invariant surface (as it is the case at zero
forcing when the attractor is dense on the full energy
surface).\annota{This condition will be verified automatically at
small forcing , if true at zero forrrcing, as a reflection of a
property called ``{\it structural stability}'' of chaotic (\ie
transitive Anosov) systems. It is important that ``small forcing''
does not mean infinetismal forcing but just not too large forcing, so
that we are really out of equilibrium. How far out of equilibrium will
depend on the model: in simple models it turns out, experimentally, to
be a property that holds for very strong forcing in the relevant
physical units.}

It is interesting to remark that at least in the limit case in which
the forcing tends to $0$, hence $\s_+\to0$, the relation (\ref{(4.3)})
becomes degenerate, but by dividing both sides by the appropriate
powers of the external fields {\it one gets a meaningful nontrivial
limit} which just tells us that Green--Kubo relations and Onsager
reciprocity hold so that (\ref{(4.3)}) can be considered an extension of
such relations, [Ga96b].

In nonequilibrium, unlike in the equilibrium case, we do not have any
well established nonequilibrium thermodynamics, but at least we have
that the (\ref{(4.3)}) is an extension valid in ``great generality''
which coincides with the only universally accepted nonequilibrium
thermodynamics relations known at $0$ forcing.

\end{section}

\begin{section}{Reversible versus irreversible thermostats.}

One would like more: it would be nice that (\ref{(4.3)}) could be
regarded as a general theorem also valid for systems which are not
time reversible. In fact there are many cases, in particular in the
theory of fluids, in which the thermostating effects are modeled by
``irreversible'' forces like friction, viscosity, resistivity, \etc

This is an important problem and it deserves further analysis: a
proposal which has been advanced, [Ga96a], is that one can imagine to
thermostat a system in various ways which are ``physically
equivalent'' (\eg one can use different models of thermostating
forces). This means that the stationary state distribution $\m$ that
describes the statistics of the system will depend on the special
equations of motions used (which reflect a particular thermostating
mechanism). However by ``suitably'', see[Ga97a], [Ga99b], changing the
equations the averages of the interesting observables will not change,
at least if the number of degrees of systems is large (\ie ``in the
thermodynamic limit'').

And it is possible  that the {\it same system} can be equivalently
described in terms of reversible or of irreversible equations of
motion.  This may at the beginning be very surprising: but in my
view it points at one of the more promising directions of research in
nonequilibrium: if correct this means that in various cases we could
use reversible equations to describe phenomena typically described by
irreversible equations: at least as far as the evaluation of several
averages is concerned. This is very close to what we are used to, since
the classical (although scarcely quoted and scarcely known) paper of
Boltzmann [Bo84] where we learned that one could use {\it different
equilibrium ensembles} to describe the {\it same system}: canonical or
microcanonical ensembles give the same average to all ``local
observables''.

In equilibrium the ensembles are characterized in terms of a few
parameters (in the microcanonical ensemble one fixes the total energy
and the specific volume, in the canonical ensemble one fixes the
temperature and the specific volume, \etc): the distributions in phase
space that correspond to the elements of such ensembles are very
different. However, if the parameters that characterize the distribution
are correctly correspondent, then the distributions give the same
averages to large classes of observables.  

Philosophically this is rather daring and physically it seems to bear
possibly important consequences: namely it might be that the general
results that apply to reversible systems do apply as well to
irreversible systems because the latter may be just equivalent to
reversible ones. We again fall dangerously close to the paradoxes,
that were used to counter the new equilibrium statistical mechanics,
by Loschmidt and Zermelo, [Ce99]. Except that now we have learnt,
after Boltzmann, why they may be in fact circumvented, [Le93].

Of course an idea like the one above has to be supported by some further
evidence and requires further investigations. I think that some rather
strong evidence in favor of it, besides its fascination, is that a
number of experiments on computer simulations of fluids motion in
turbulent states have been already carried out (for other purposes) but
their results can be interpreted as evidence in favor of the new
idea, [SJ93]: see [Ga97a], [Ga97b], [Ga99b].

More recently there have been attempts to perform {\it dedicated
simulations} to observe this property in fluid dynamics systems.
These experiments are interesting also because the application of the
chaotic hypothesis to fluids may look less direct than to nonequilibrium
statistical systems. In fact trying to perform experiments, even
in simulations, is quite promising and perhaps we may even be close to
the possibility of critical tests of the chaotic hypothesis in real fluids.
The preliminary results of simulations are encouraging but more work
will have to be done, [RS98]; in the next section we shall examine a
real experiment and attempt a theoretical explanation of it.

\end{section}

\begin{section}{An experiment with water in a Couette flow.}

A most interesting experiment by Ciliberto--Laroche, [CL98], on a
physically macroscopic system (water in a container of a size of the
order of a few deciliters), has been performed with the aim of testing
the relation (\ref{(4.3)}). 

This being a real experiment one has to stretch quite a bit the very
primitive theory developed so far in order to interpret it and one has
to add to the chaotic hypothesis other assumptions that have been
discussed in [BG97], [Ga97a].

The experiment attempts at measuring a quantity that is eventually
interpreted as the difference $\z(p)-\z(-p)$, by observing the
fluctuations of the product $\th u^z$ where $\th$ is the {\it
deviation} of the temperature from the average temperature in a small
volume element $\D$ of water at a fixed position in a Couette flow and
$u^z$ is the velocity in the $z$ direction of the water in the same
volume element.

The result of the experiment is in a way quite unexpected: it is found
that the function $\z(p)$ is rather irregular and lacking symmetry
around $p=1$ {\it but the function} $\z(p)-\z(-p)$ {\it seems to be
strikingly linear}. As discussed in [Ga97a], predicting the slope of
the entropy creation rate would be difficult but if the equivalence
conjecture considered above and discussed more in detail in [Ga97a] is
correct then we should expect linearity of $\z(p)-\z(-p)$. In the
experiment of [CL98] the quantity $\th u_z$ does not appear to be the
divergence of the phase space volume simply because there is no model
proposed for the theory of the experiment. Nevertheless
Ciliberto--Laroche select the quantity $\int_\D \th\,u^z \,d\V x$ on
the basis of considerations on entropy and dissipation so
that there is great hope that in a model of the flow this quantity can
be related to the entropy creation rate discussed in \S3.

Here we propose that a model for the equations, that can be reasonably
used, is Rayleigh's model of convection, [Lo63], [Ga97b] sec. 5. An
attempt for a theory of the experiment could be the following.

One supposes that the equations of motion of the system in the whole
container are written for the quantities $t,x,z,\th,u$ in terms of the
height $H$ of the container (assumed to be a horizontal infinite
layer), of the temperature difference between top and bottom $\d T$
and in terms of the phenomenological ``friction constants'' $\n,\ch$
of viscosity, dynamical thermal conductivity and of the thermodynamic
dilatation coefficient $\a$. We suppose that the fluid is
$3$--dimensional but stratified, so that velocity and temperature
fields do not depend on the coordinate $y$, and gravity is directed
along the $z$--axis: $\V g= g\, \V e,\, \V e=(0,0,-1)$.

In such conditions the equations, including the boundary conditions
(of fixed temperature at top and bottom and zero normal velocity at
top and bottom), the convection equations in the Rayleigh model, see
[Lo63] eq. (17), (18) where they are called the {\it Saltzman
equations}, and [Ga97b], become
\begin{eqnarray}
&\Dpr\cdot\V u=0\nonumber\\ &\dot{\V u}+\W u\cdot\W \partial\,\V u=\n\D\V
u-\a\th\V g-\Dpr p'\nonumber\\ &\dot\th+\W
u\cdot\W\partial\th=\chi\D\th+{\d T\over H} u_z\\ &\th(0)=0=\th(H),\quad
u_z(0)=0=u_z(H),\nonumber\\ 
&\int u_x d\V{x}=\int u_yd\V{x}=0
\label{(6.1)}\end{eqnarray}
The function $p'$ is related, {\it but not equal}, to the pressure
$p$: within the approximations it is $p=p_0-\r_0 g z+ p'$.

It is useful to define the following adimensional quantities
\begin{eqnarray}
\t&=&t\n H^{-2},\ \x= x H^{-1},\ \h=y H^{-1},\ \z=z H^{-1},\nonumber\\
\th^0&=&\frac{\a\th}{\a\,\d T},\ \V u^0=(\sqrt{g H\a\,\d T})^{-1}\,\V
u\nonumber\\ 
R^2&=&{g H^3\a\,\d T\over\n^2},\qquad R_{Pr}={\n\over\k}
\label{(6.2)}\end{eqnarray}
and one checks that the Rayleigh equations take the form
\begin{eqnarray}
&\dot\V u+ R\W u\cdot\W\partial\,\V u=\D\V u- R\th\V e-\Dpr p,\nonumber\\
&\dot\th+ R\W u\cdot\W\partial\,\th=R_{Pr}^{-1}\D\th+ R u_z,\nonumber\\
&\Dpr\cdot\V u=0\nonumber\\
&u_z(0)=u_z(1)=0,\qquad\th(0)=\th(1)=0,\nonumber\\
&\int u_x d\V{x}=\int u_y
d\V{x}=0\label{(6.3)}\end{eqnarray}
where we again call $t,x,y,z,\V u,\th$ the adimensional coordinates
$\t,\x,\h,\z, \V u^0,$ $\th^0$ in (\ref{(6.2)}). The numbers $R, R_{Pr}$ are
respectively called the {\it Reynolds and Prandtl numbers} of the
problem: $R_{Pr}=\sim6.7$ for water while $R$ is a parameter that we
can adjust, to some extent, from $0$ up to a rather large value.

According to the principle of equivalence stated in [Ga97a] here one
should impose the constraints
\begin{equation}
\int \big( \,(\Dpr \W u)^2\,+\,\frac1{R_{Pr}}
(\Dpr \th)^2\,\big)\,d\V x= C\label{(6.4)}\end{equation}
on the ``frictionless equations'', (\ie (\ref{(6.3)}) without the terms
with the laplacians) obtaining
\begin{eqnarray}
&\Dpr\cdot\V u=0\nonumber\\
&\dot{\V u}+R\,\W u\cdot\W \partial\,\V u=R\th\,\V e-\Dpr p'+ 
\V\t_{th}
\nonumber\\
&\dot\th+\W u\cdot\W\partial\th=Ru^z+\l_{th}\nonumber\\
&\th(0)=0=\th(H),\quad \int u_x d\V{x}=\int u_yd\V{x}=0
\label{(6.5)}\end{eqnarray}
where the frictionless equations are modified by 
{\it the thermostats forces} $\V\t_{th},\l_{th}$: the latter impose
the nonholonomic constraint in (\ref{(6.4)}). Looking only at the bulk
terms we see that the equations obtained by imposing the constraints
via Gauss' principle, see [Ga96a], [Ga97a], become the (\ref{(6.3)}) with
coefficients in front of the Laplace operators equal to $\n_G,\n_G
R_{Pr}^{-1}$, respectively, with the ``gaussian multiplier'' $\n_G$
being an {\it odd} functions of $\V u$, see [Ga97a]: setting
$\tilde C=\int \big((\D \V u)^2+R_{Pr}^{-1}(\D \th)^2\big) d\V x$ one
finds
\begin{eqnarray}
\n_G=&\tilde C^{-1}\Big(\int\big((\D\V u\cdot(\W u\cdot\W\partial)\V
u)+\nonumber\\
&+R_{Pr}^{-2} (\D\th\cdot(\W u\cdot\W\partial)\th)+\nonumber\\
&+R(1+R_{Pr}^{-1})
\,u^z\th\big)\,\,d\V x\Big)\label{(6.6)}\end{eqnarray}
which we write $\n_G=\n^i+R \n^e$. And the equations become, finally

\begin{eqnarray}
&\Dpr\cdot\V u=0\nonumber\\
&\dot{\V u}+R\,\W u\cdot\W \partial\,\V u=R\th\,\V e-\Dpr p'+\n_G \D\V
u\nonumber\\
&\dot\th+\W u\cdot\W\partial\th=Ru^z+\n_G\frac1{R_{Pr}}\D \th\nonumber\\
&\th(0)=0=\th(H),\quad \int u_x d\V{x}=\int u_yd\V{x}=0\label{(6.7)}
\end{eqnarray}
One has to tune, [Ga97a], the value of the constant $C$ in
(\ref{(6.4)}) so that the average value of $\n_G$ is precisely the
physical one: namely $\media{\n_G}=1$ by (\ref{(6.3)}). This is the
same, in spirit, as fixing the temperature in the canonical ensemble
so that it agrees with the microcanonical temperature thus implying
that the two ensembles give the same averages to the local
observables.

The equations (\ref{(6.7)}) are time reversible (unlike the (\ref{(6.3)}))
under the time reversal map:

\begin{equation}I(\V u,\th)=(-\V u,\th)\label{(6.8)}\end{equation}
and they should be supposed, by the arguments in [Ga97a], ``equivalent''
to the irreversible ones (\ref{(6.3)}).

The (\ref{(6.7)}) should have, by the general theory of [Ga97a], a
``divergence'' $\s(\V u,\th)$ whose fluctuation function $\z(p)$
verifies a linear fluctuation relation, \ie $\z(p)-\z(-p)$ should be
linear in $p$ similar to (\ref{(4.3)}). And the divergence of the
above equations is proportional to $\n_G$ if one supposes that the
high momenta modes can be set equal to $0$ so that the equation
(\ref{(6.7)}) becomes a system of finite differential equations for
the Fourier components of $\V u,\th$. The Lorenz' equations, for
instance, reduced the number of Fourier components necessary to
describe (\ref{(6.3)}) to just three components, thus turning it into
a system of three differential equations.

By the conjectures in [Ga97a] a fluctuation relation should hold for
the divergence; except that the slope of the differenxe $\z(p)-\z(-p)$
should not necessarily be $\s_+$ as in (\ref{(3.3)}). 

Proceeding in this way the divergence of the equations of motion is a
sum of two integrals one of which proportional to the Reynolds number
$R$.  If instead of integrating over the whole sample we integrate
over a small region $\D$, like in the experiment of [CL98], we can
expect to see a fluctuation relation for the entropy creation rate
only if the fluctuation theorem {\it holds locally}, \ie for the
entropy creation in a small region.

This is certainly not implied by the proof in [GC95]: however {\it
when the dissipation is homogeneous through the system}, as it is the
case in the Rayleigh model there is hope
that the fluctuation relation holds locally, again for the same
reasons behind the equivalence conjecture (\ie a small subsystem
should be equivalent to a large one). The actual possibility of a
local fluctuation theorem in systems with homogeneous dissipation has
been shown in [Ga99a], after having been found through numerical
simulations in [PG99], and therefore we can imagine that it might
apply to the present situation as well.

If the contributions to the entropy creation due to the term $R\int_\D
u_z \th\,d\V x$, where $\D$ is the region where the measurements of
[CL98] are performed, dominate over the others we have an explanation
of the remarkable experimental result. Unfortunately in the experiment
[CL98] the contributions not explicitly proportional to $R$ to the
entropy creation rates have not been measured. But the Authors hint
that they should indeed be smaller; in any event they might be
measurable by improving the same apparatus, so that one can check
whether the above attempt to an explanation of the experiment is
correct, or try to find out more about the theory in case it is not
right. If correct the above ``theory'' the experiment in [CL98] would
be quite important for the status of the chaotic hypothesis.

\end{section}
\begin{section}{Conclusions.}

We have tried to show how, still today, one can attribute to the motions of
complex systems the character of periodic motions, as in the observation
in [Bo66] which gave birth to modern statistical mechanics of
equilibrium. Yet such periodic motions are motions of huge period and
they cannot be confused with the epicyclical motions of Aristoteles
which survived in mechanics until Boltzmann and Poincar\'e and in fluid
mechanics until the early 1960's.

Thinking all motions as periodic allows us to unify the statistical
mechanics of equilibrium and nonequilibrium and at the same time to
unify them with the theory of (developed) turbulence. It also shows
that the discrete viewpoint of Boltzmann which started with an attempt
to save the Aristotelian view of motion (as in the quoted passage of
[Bo66]), and which is {\it necessary} to avoid contradictions inherent
in the dogmatic conception of space time as a continuum, is very
powerful also to attack problems that seemed treatable {\it only} by
very refined mathematical analysis.

And in nonequilibrium problems a theory of {\it statistical ensembles}
might be possible that extends in a bold and surprising way the theory
of the equilibrium ensembles: in this theory the phenomenological
constants that appear in the equations of motion of thermostated
systems can be replaced by fluctuating quantities {\it with
appropriate averages} turning certain {\it other} fluctuating
observables into {\it exact constants}. This is ``as'' in equilibrium
where we can introduce a {\it constant} canonical temperature by
imposing that it is equal to the {\it average} of the fluctuating
microcanonical temperature (\ie kinetic energy). In this way we are
not forced to attribute a fundamental role to the phenomenological
transport coefficients: they are just convenient Lagrange multipliers
for the statistics of the stationary states. Like the temperature o
the activity in equilibrium statistical mechanics of the canonical or
grand canonical ensembles.

The fluctuation theorem seems to open the way to considerations over
out-of-equilibrium systems that were unthinkable until recently:
perhaps this is not the right approach but it has led to interesting
experimental questions which might attract more interest in the
future. 

The above picture is lacking sufficient experimental confirmations to
be considered established or even likely: it has to be regarded at the
moment as one more attempt among many in this century to understand a
difficult problem.  We should not forget that the whole XX-th century
failed to give us a theory of nonequilibrium phenomena and of
turbulence which could be regarded as fundamental as the
Boltzmann--Maxwell--Gibbs principles of equilibrium statistical
mechanics: the problem is so fundamental that it will (almost)
certainly attract the attention of the new generations of physicists
and a solution of it is certainly awaited.

\end{section}
\*
{\it Acknowledgements: work supported by CNR, Consiglio Nazionale
delle Ricerche, the Italian Science Council.}

\small
\*
\0{\bf Bibliography:}
\*

\0[AA89] {\it The astronomical Almanac for the year 1989}, issued by the
National Almanac Office, US government printing office, 1989.

\0[An82] Andrej, L.: {\it The rate of entropy change in
non--Hamiltonian systems}, Physics Letters, {\bf 111A}, 45--46, 1982.

\0[BG97] Bonetto, F., Gallavotti, G.: {\it Reversibility, coarse 
graining and the chaoticity principle}, Communications in Mathematical
Physics, {\bf189}, 263--276, 1997.

\0[Bo66] Boltzmann, L.: {\it\"Uber die mechanische Bedeutung des zweiten
Haupsatzes der W\"armetheorie}, in "Wissenschaftliche Abhandlungen",
ed. F. Ha\-sen\-\"ohrl, vol. I, p. 9--33, reprinted by Chelsea, New York.

\0[Bo74] Boltzmann, L.: {\it Theoretical Physics and philosophical
writings}, ed. B. Mc Guinness, Reidel, Dordrecht, 1974.

\0[Bo84] Boltzmann, L.: {\it \"Uber die Eigenshaften monzyklischer
und anderer damit verwandter Systeme}, in "Wissenshafltliche
Abhandlungen", ed. F.P. Hasen\"ohrl, vol. III, p. 122--152,
Chelsea, New York, 1968, (reprint).

\0[Bo95] Boltzmann, L.: {\it Lectures on gas theory}, Dover, 1995 (reprint).

\0[CG99] Cohen, E.G.D., Gallavotti, G.: {\it Note on Two Theorems in 
Nonequilibrium Statistical Mechanics}, archived in mp$\_$arc 99-88,
cond-mat 9903418, in print in J. Stat. Phys.

\0[CL98] Ciliberto, S., Laroche, C.: {\it An experimental
verification of the Galla\-vot\-ti--Cohen fluctuation theorem}, Journal de
Physique, {\bf8}, 215--222, 1998.

\0[Ce99] Cercignani, C.: {\sl Ludwig Boltzmann}, Oxford, 1998.

\0[Co30] Copernicus, N.: {\it De hypothesibus motuum
caelestium a se constitutis commentariolus}, in {\sl Opere},
ed. L. Geymonat, UTET, Torino, 1979.

\0[ECM93] Evans, D.J.,Cohen, E.G.D., Morriss, G.P.: {\it Probability
of second law violations in shearing steady flows}, Physical Review
Letters, {\bf 71}, 2401--2404, 1993.

\0[EE11] Ehrenfest, P., Ehrenfest, T.: {\it The conceptual
foundations of the statistical approach in Mechanics},
Dover, New York, 1990, (reprint).

\0[EPR98] Eckmann, J.P., Pillet, A., Rey--Bellet, L.: {\it
Non-Equilibrium Statistical Mechanics of Anharmonic Chains\\ Coupled to
Two Heat Baths at Different Temperatures}, in mp$\_$arc \# 98-233.

\0[GC95] Gallavotti, G., Cohen, E.G.D.: {\it Dynamical
ensembles in non-equili\-bri\-um statistical mechanics}, Physical Review
Letters, {\bf74}, 2694--2697, 1995. Galla\-vot\-ti, G., Cohen,
E.G.D.: {\it Dynamical ensembles in stationary states}, Journal of
Statistical Physics, {\bf 80}, 931--970, 1995.

\0[Ga65] Galilei, G.: {\it Il saggiatore}, edited by L. Sosio,
Feltrinelli, Milano, 1965 (reprint).

\0[Ga95]  Gallavotti, G.: {\it Ergodicity, ensembles, irreversibility
in Boltzmann and beyond}, Journal of Statistical Physics, {\bf 78},
1571--1589, 1995.

\0[Ga96a]  Gallavotti, G.: {\it Chaotic hypothesis: Onsager reciprocity and 
   fluc\-tua\-tion--dis\-si\-pa\-tion theorem}, Journal of Statistical Physics,
   {\bf 84}, 899--926, 1996.

\0[Ga96b] Gallavotti, G.: {\it Extension of Onsager's reciprocity to
   large fields and the chao\-tic hypothesis}, Physical Review
   Letters, {\bf 78}, 4334--4337, 1996.

\0[Ga97a]  Gallavotti, G.: {\it Dynamical ensembles equivalence 
in fluid mechanics}, Physica D, {\bf 105}, 163--184, 1997.

\0[Ga97b] Gallavotti, G.: {\it Ipotesi per una introduzione alla
Meccanica dei Fluidi}, ``Qua\-der\-ni del CNR-GNFM'', vol. {\bf },
p. 1--428, Firenze, 1997. English translation in preparation.

\0[Ga99a] Gallavotti, G.: {\it A local fluctuation theorem}, Physica A,
   {\bf 263}, 39--50, 1999.

\0[Ga99b] Gallavotti, G.: {\sl Statistical mechanics}, Springer Verlag,
1999.

\0[LL67] Landau, L., Lifschitz, L.E.: {\sl M\'ecanique des fluides}
MIR, Moscow, 1971.

\0[La74] Lanford, O.: {\it Time evolution of large classical systems},
in ``Dynamical systems, theory and applications'', p. 1--111, ed. J.
Moser, Lecture Notes in Physics, vol. 38, Springer Verlag, Berlin, 1974.

\0[Le93] Lebowitz, J.L.: {\it Boltzmann's entropy and time's
arrow}, Physics Today, 32--38, 1993.

\0[Lo63] Lorenz, E.: {\it Deterministic non periodic flow}, J. 
of the Atmospheric Sciences, 20, 130- 141, 1963.

\0[PG99] Perroni, F., Gallavotti, G.: {\it An experimental test of the
local fluctuation theorem in chains of weakly interacting Anosov
systems}, preprint, 1999, in http://ipparco. roma1. infn. it at the 1999
page.

\0[PH92] Posh, H.A., Hoover, W.G.: Phys.Rev., {\bf A38}, 473,
1988; in ``Molecular Liquids: new perspectives'', ed. Teixeira-Dias,
Kluwer, 1992.

\0[Po87] Poincar\'e, H.: {\sl Les m\'ethodes nouvelles de la
m\'ecanique c\'eleste}, original 1983, reprint by A. Blanchard, Paris,
1987, tome {\bf II}, Ch. IX and XI.
 
\0[RS98] Rondoni, L., Segre, E.: {\it Fluctuations in 2D reversibly-damped
turbulence}, in chao-dyn/9810028.

\0[RT71a] Ruelle, D., Takens, F.: {\it On the nature of 
turbulence}, Communications in mathematical Physics, {\bf20}, 167,
1971.  

\0[RT71b] Ruelle, D., Takens, F.: {\it Note concerning our paper "On the
nature of turbulence"}, Communications in Mathematical Physics, {\bf
23}, 343--344, 1971.

\0[Ru78] Ruelle, D.: {\it What are the measures describing turbulence?},
Progress of Theoretical Physics, (Supplement) {\bf64}, 339--345,
1978. See also {\it Measures describing a turbulent flow}, Annals of
the New York Academy of Sciences, volume {\sl Nonlinear dynamics},
ed. R.H.G. Helleman, {\bf357}, 1--9, 1980.

\0[Ru96] Ruelle, D.: {\it Positivity of entropy production in
non\-equi\-li\-brium statistical mechanics}, Journal of Statistical Phy\-sics,
{\bf 85}, 1--25, 1996.

\0[Ru98] Russo, L.: {\it The definitions of fundamental geometric
entities contained in book I of Euclids Elements}, Archive for the
History of Exact Sciences, {\bf 52}, 195--219, 1998.

\0[Ru99] Ruelle, D.: {\it Smooth dynamics and new theoretical ideas in
non-equi\-li\-brium statistical mechanics}, Lecture notes, Rutgers
University, mp$\_$arc \#98-770, 1998, or chao-dyn \#9812032, in print on
Journal of Statistical Physics.

\0[SJ93] She Z.S., Jackson, E.: {\it Constrained Euler system for
Navier--Stokes turbulence}, Physical Review Letters, {\bf 70},
1255--1258, 1993.

\0[Si77] Sinai, Y.G.: {\sl Lectures in ergodic theory}, Lecture notes
in Mathematics, Prin\-ce\-ton U. Press, Princeton, 1977.

\0[Si79] Sinai, Y.G.: {\it Development of Krylov's ideas},
in Krylov, N.S.: {\sl Works on the foundations in statistical
physics}, Princeton University Press, 1979, p. 239--281.

\0[Sm67] Smale, S.: {\it Differentiable dynamical
systems}, Bullettin of the American Mathematical Society, {\bf 73 },
747--818, 1967.

\*
\0{\it Giovanni Gallavotti

\0Dipartimento di Fisica

\0Universit\'a di Roma ``La Sapienza''

\0P.le Moro 2, 00185 Roma, Italy

\0email: { giovanni@ipparco.roma1.infn.it}

\0tel. +39-06-49914370, fax. +39-06-4957697}
\*

\def\revtex{R\raise2pt\hbox{E}VT\lower2pt\hbox{E}X}
\0(slightly corrupted) \revtex

\ciao

          Lucio Russo:
          The Definitions of Fundamental Geometric Entities Contained in Book I of Euclids Elements
          Arch Hist Exact Sc. 52 (1998) 3, 195-219
          A